\documentclass[a4paper,aps,pra,showpacs,preprintnumbers]{revtex4}
\usepackage{graphics,graphicx,epsfig}
\usepackage[centertags]{amsmath}
\usepackage{amsfonts}
\usepackage{amssymb}
\usepackage[english]{babel}
\usepackage{graphicx}
\usepackage{amsthm,color}
\usepackage{ulem}                   

\begin{document}

\def\BE{\begin{equation}}
\def\EE{\end{equation}}
\def\BEA{\begin{eqnarray}}
\def\EEA{\end{eqnarray}}
\def\BY{\begin{eqnarray}}
\def\EY{\end{eqnarray}}

\def\L{\label}
\def\nn{\nonumber}
\def\ds{\displaystyle}
\def\o{\overline}

\def\({\left (}
\def\){\right )}
\def\[{\left [}
\def\]{\right]}
\def\<{\langle}
\def\>{\rangle}

\def\h{\hat}
\def\hs{\hat{\sigma}}
\def\td{\tilde}

\def\k{\mathbf{k}}
\def\q{\mathbf{q}}
\def\r{\vec{r}}
\def\ro{\vec{\rho}}
\def\a{\hat{a}}
\def\b{\hat{b}}
\def\c{\hat{c}}
\def\h{\hat}

\title{Manipulation of quantum states in memory cell: controllable Mach-Zehnder interferometer} \vspace{1cm}
\author{ A.S.~Losev, T.Yu.~Golubeva, Yu.M.~Golubev}
\address{St.Petersburg State University, 7/9 Universitetskaya nab., St. Petersburg, 199034 Russia}
\date{\today}

\begin{abstract}
In this article, we consider the possibility of manipulation of quantum signals, ensured by the use of the tripod-type atomic memory cell. We show
that depending on a configuration of driving fields at the writing and reading, such a cell allows both to store and transform the signal. It is
possible to provide the operation of the memory cell in a Mach-Zehnder interferometer mode passing two successive pulses at the input. We proposed a
procedure of partial signal read out that provides entanglement between the retrieved light and the atomic ensemble. Thus, we have shown that tripod
atomic cell is a promising candidate to implement of quantum logical operations, including two-qubit ones, that can be performed on the basis of only
one cell.

\end{abstract}
 \maketitle

\section{Introduction\L{I}}

Nowadays, quantum memory cells becomes considered not only as elements of quantum communication networks reproducing an original signal on demand
(see, e.g., reviews \cite{Sangouard-2011, Hammerer-2010, Simon-2010, Lvovsky-2009}) but also as signal converters (quantum transistors), which can be
used in photonic quantum computation systems \cite{Humphreys-2014}. The experiments, where a quantum memory is utilized to generate a random secret
key by modifying the read signal \cite{Su-2016} , or to encode and decode an information in quantum secure direct communication protocols
\cite{Zhang-2016} were implemented. The protocols for realization of the quantum random access memory \cite{Moiseev-2016} and the schemes of
enhancing multiphoton rates with quantum memories \cite{Nunn-2013} were offered. The memory cell working as a coherent and dynamic beam splitter was
demonstrated experimentally \cite{Kim-2016}. The use of the memory cell as the coherent beam splitter allows to employ it in the photonic computation
systems to perform single-qubit logic operations. However, as is well known from the information theory, one have to realize not only single-qubit
but also two-qubit procedures such as, for example, C-Phase and C-Not, to implement a full set of logic operations.

Performing of two-qubit photonic operations is hampered by the weakness of photon-photon interactions. The number of proposals are offered to get
over this difficulty. In particular, schemes of temporal \cite{Leuchs-2012} and spatial \cite{Sondermann-2007} light mode shaping are developed. They
provide an effective control of a qubit state. There is an idea of quantum states swaping between two separated atomic ensembles via a single atom
playing the role of controllable gate \cite{Moiseev-2013}.

We discuss here the additional possibility of manipulation of quantum signals, ensured by the use of the tripod-type atomic memory cell. We show that
such cell allows not only to store the signal, but also to transform it depending on a configuration of driving fields at the writing and reading. In
contrast to the $\Lambda$-type light-matter interaction, where the only one spin wave can be excited, in tripod memory cells one can provide the
excitation of two spin waves. Moreover, depending on a selected basis of states, these spin waves can be statistically independent or correlated
(e.g., entangled). Recent experiments with ensembles of atoms in the tripod configuration showed the controllable splitting and modulation of
single-photon-level pulses \cite{Yang-2015}. It is also shown that the implementation of the interference measurements based on the double-tripod
scheme \cite{Lee-2014} is possible.

The most exciting option of the application of the tripod-type atomic cell is the ability to organize its interaction with two successive input
signals. Herewith, choosing a particular configuration of the read out driving fields, we can, for example, convert two input orthogonally squeezed
signal pulses into two entangled pulses at the output of the cell, or to entangle the output light with a spin wave. We show that it is possible to
provide the work of the quantum memory in a regime of the interferometer used only one single tripod scheme, in contrast to authors of the work
\cite{Lee-2014}. For this purpose, we exploit the possibility to excite the medium successively with the help of various combinations of driving
fields.

Note, that a light-matter entanglement was demonstrated recently experimentally in scheme of four-level photon echo-type quantum memory using the
process of rephased amplified spontaneous emission in doped crystals \cite{Ferguson-2016}. The creation of such states provides a unique opportunity
to transform the modes of the medium by acting on the read light.

Continuing the analogy of the authors \cite{Kim-2016}, discussing the work of the $\Lambda$-type memory cell as the beam splitter, we say that the
tripod-type memory cell is similar to the Mach-Zehnder interferometer. Such enrichment of the quantum memory protocol opens up additional
possibilities for its application not only to store and to retrieve on demand but also to manipulate the stored states within the memory cell.

In this article, we consider the so-called "high-speed resonance"\ memory \cite{Golubeva-2011} as a mechanism of the light-matter interaction. This
type of interaction is realized when the pulse durations of the signal and driving fields are much shorter than the lifetime of the excited state.
This interaction takes less time in comparison with the adiabatic or EIT memory, and we consider it more prospective for the purposes of quantum
computation. Nevertheless, the results of this study can be easily transferred to the other types of the interaction (EIT, adiabatic or Raman
memory).

The article is structured as follows. In Section \ref{II}, the model under consideration is presented and the Heisenberg equations for broadband
memory in tripod-type atomic ensemble are derived. Section \ref{III} is devoted to the discussion of the solutions in two different bases employing
for spin wave description. In Section \ref{IV} we consider a modification of the Schmidt decomposition for separate mode analysis of the writing and
reading stages. In the \ref{V}th Section we study the modes structure of the input signal mapped into the memory cell. Squeezed pulses of light from
the synchronized sub-Poissonian laser are considered as a source of the input signal. The \ref{VI}th Section is devoted to the quantum statistical
properties of spin waves after the writing stage as well as of the retrieved light after the full memory cycle in that case, when only one pulse is
supplied to the input cell. In addition, here the possibility to generate the combined (semi-squeezed and semi-entangled) quantum state (CQS) of
spins and light is demonstrated. Finally, in Section \ref{VII} we consider the generation of different states of the medium and light, when two
orthogonally squeezed pulses are passed on the input of the cell. In particular, we discuss a way to generate the entangled light-matter state.

\section{Main equations of broadband resonant tripod memory \L{II}}

In this Section we will apply a formal approach for the broadband resonant quantum memory in the tripod-type atomic ensemble, which was discussed in
details in the article \cite{Losev-2016}. The actual energy structure of each atom and electromagnetic fields is presented in Fig. \ref{Fig1}. It is
assumed that initially all atoms are pumped into the ground state on sublevel $|3\rangle$. Mapping the non-classical signal field with a slowly
varying Heisenberg amplitude $\a$ is governed by the operation of this field on the transition $|3\>-|4\>$ accompanied by the action of two driving
fields on the transitions $|1\>-|4\>$ and $|2\>-|4\>$. We assume that the carrier frequencies of all fields coincide with ones of the corresponding
atomic transitions (on-resonant configuration). The writing process is determined by the appearance of two atomic coherences on the transitions $|1\>
- |3\>$ and $|2\> - |3\>$.
\begin{figure}[h]
\centering
\includegraphics[height=34mm]{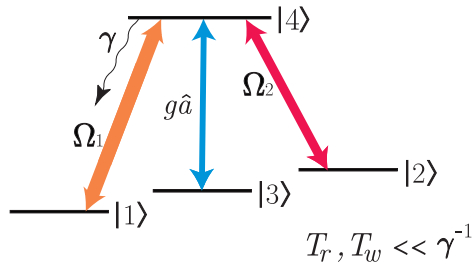}
\caption{Tripod type atomic configuration, $\Omega_1$ and $\Omega_2$ are Rabi frequencies of the driving fields, $\h{a}$ is the slowly varying
Heisenberg amplitude of the signal field.} \L{Fig1}
\end{figure}
In the article \cite{Losev-2016} a set of Heisenberg differential equations for spatial-time evolution of the slowly varying amplitude of the signal
field $\a(t,z)$ and three spin waves $\b_1(t,z), \b_2(t,z)$ and $\c(t,z)$ is obtained:
\BY
&&\partial_z \a(t,z) = -g \sqrt{N} \; \h{c}(t,z), \L{1}\\
&&\partial_t \h{c}(t,z) = g \sqrt{N} \; \a(t,z) + \Omega_1 \b_1(t,z) + \Omega_2 \b_2(t,z), \L{2}\\
&&\partial_t \b_1(t,z) = -\Omega_1 \h{c}(t,z), \L{3}\\
&&\partial_t \b_2(t,z) = -\Omega_2 \h{c}(t,z). \L{4}
\EY
We applied here the plane waves approximation and neglected by the relaxation processes assuming that the times of the interaction are very short
(the broadband memory process), and the spontaneous relaxation has no time to engage. Here, $N$ is linear atomic concentration with a dimension
$cm^{-1}$, $\Omega_1$ and $\Omega_2$ are the Rabi frequencies of the driving classical fields. The constant $g$ defines the dipole interaction of the
signal and the single atom:
\BE g = \( \frac{\omega_s}{2 \varepsilon_0 \hbar c} \)^{\!1/2} \!\!\!\! d_{43}, \L{5} \EE
where $\omega_s$ is the frequency of the signal field, $d_{43}$ is a dipole moment of the transition $|3\>-|4\>$.

The Heisenberg amplitude of the signal field $\a(t,z)$ obeys the canonical commutation relations:
\BEA &&\[ \a(z,t), \a^\dag(z,t^{\prime}) \] = \delta (t - t^{\prime}),\qquad \[ \a(z,t), \a^\dag(z^{\prime},t) \] = c \( 1-\frac{i}{k_s}
\partial_z \) \delta (z - z^{\prime}). \L{5}
 \L{6} \EEA

The evolution of the atoms is defined by the collective Heisenberg amplitudes $\b_1$, $\b_2$ and $\c$. We assume that the population of the level
$|3\>$ is much higher than the number of the signal photons, that impose this population to be practically unchangeable. The operators of the
collective spin waves can be expressed via the coherences of the single atoms $\hs^a_{13}$, $\hs^a_{32}$, $\hs^a_{34}$ as follows:
\BY
&& \hat b_1(t,z)=\frac{1}{\sqrt N}\sum_a \hat\sigma^a_{13}(t)\delta(z-z_a),\qquad\hat b_2(t,z)=\frac{1}{\sqrt N}\sum_a
\hat\sigma^a_{32}(t)\delta(z-z_a),\qquad\hat c(t,z)=\frac{1}{\sqrt N}\sum_a
\hat\sigma^a_{34}(t)\delta(z-z_a).\nn\\
\EY
Here, a summation is over the all atoms of the ensemble. Then, the collective spin operators obey the following commutation relations:
\BY
&&\[\hat b_i(t,z),\hat b^\dag_j(t,z^\prime)\]=\delta_{ij}\delta(z-z^\prime)\quad (i,j=1,2),\qquad\[\hat c(t,z),\hat c^\dag
(t,z^\prime)\]=\delta(z-z^\prime).
\EY
%

\section{Writing and reading out of the signal \L{III}}

According to the equations (\ref{1})-(\ref{4}), mapping the signal pulse into the medium and further retrieval of the signal depend on a ratio
between driving fields, that is, depend on the values of Rabi frequencies $\Omega_1$ and $\Omega_2$. For example, if one of these values is equal to
zero, then only one of the spin-waves, $\hat b_{1}$ or $\hat b_{2}$, can be excited in the medium (or can be read out from the medium). At the same
time, the possibility to write the second signal pulse into the second channel survives.

The opportunities of manipulation become even more expanded, when we take into account the following. Let us consider the interaction under condition
$\Omega_1=\pm\Omega_2$, when both spins $\hat b_{1}$ and $\hat b_{2}$ are exited at the same time. Herewith, equations (\ref{1})-(\ref{4}) can be
rewritten in the form:
\BEA
&&\partial_z \a(t,z) = - g \sqrt{N}\; \c(t,z) \L{9}\\
&&\partial_t \c(t,z) = g \sqrt{N}\; \a(t,z) + \Omega \;\b_\pm(t,z) \L{10}\\
&&\partial_t \b_\pm(t,z) = -\Omega \;\h{c}(t,z) \L{11}\\
&&\partial_t \b_\mp(t,z) =  0. \L{12} \EEA
Here, we determined a variable $\Omega$ by an equality $\Omega=\Omega_1 \sqrt2=\pm\Omega_2\sqrt2$. Instead of canonical amplitudes $\hat b_1$ and
$\hat b_{2}$, the new ones have appeared here:
\BY
&&\hat b_{\pm}=\frac{1}{\sqrt2}\(\hat b_1{\pm}\hat b_{2}\).\L{9.}
\EY
Thus, speaking in terms of new spin amplitudes, only one spin wave with the amplitude $\hat b_+$ or $\hat b_{-}$ is excited.

Let us discuss four different possibilities to write the single pulse into the medium, i.e. when it excites the amplitude of one of the spin waves
$\hat b_1$, $\hat b_{2}$, $\hat b_+$ or $\hat b_{-}$.

It is easy to verify that two spin waves $\hat b_1$ and $\hat b_{2}$ (or similarly two waves $\hat b_+$ and $\hat b_{-}$) can be written into  the
medium independently of one another.

As can be seen from the equations (\ref{1})-(\ref{4}) and (\ref{9})-(\ref{12}), their structure is identical if we involve the assumption that the
following equalities are carried out: $\Omega_1=\pm\Omega_2=\Omega/\sqrt2$ for equations (\ref{9})-(\ref{12}), and $\(\Omega_1=\Omega,\;\Omega_2=0\)$
or $\(\Omega_1=0,\;\Omega_2=\Omega \)$ for equations (\ref{1})-(\ref{4}). Moreover, these equations coincide with those for the memory based on the
$\Lambda$-type atomic medium. This enables us to use the well-known formulas \cite{Golubeva-2012} and derive the required solutions in the form of
integral transformations. For example, the amplitude of the spin wave at the end of the writing process can be get in the explicit form:
\BY
&& \hat{b}_r(z)= - \frac{1}{\sqrt2} \int^{T_W\!\!\!\!\!\!}_0 dt \; \h{ a}_{in}(T_W -  t) \;G_{ab}(t,z) +\hat v_{r}(z),\qquad
r=1,2,\pm.\L{14}
\EY
Hereinafter we use the dimensionless coordinate and time according to the following transformations:
\BY
&&\Omega t\to t,\qquad {2g^2N}z/{\Omega}\to z.\L{}
\EY
Then for the duration of the writing process $T_W$ and a thickness of the medium layer $L$ we will have, respectively,
\BY
&&\Omega T_W\to T_W,\qquad{2g^2N}L/{\Omega}\to L.\L{}
\EY

In the equation (\ref{14}) and below we utilize the symbol $\hat v$ with different indexes to denote a contribution of physical subsystems, which are
initially in vacuum states. Under the calculations we can confine ourselves to comprehension of their vacuum nature. There is no need to derive the
vacuum contributions in the explicit form, because they all disappear when one calculate the normally ordered averages (exactly such averages we
should calculate further for the evaluation of the quantum memory).

The Eq. (\ref{14}) should be interpreted in the following way. When, for example, $\Omega_1=\Omega_2$ and only the spin-wave with the amplitude $\hat
b_{+}$ is excited according to (\ref{9})-(\ref{10}), we have to put $r=+$ in Eq. (\ref{14}). In regard to the wave with the amplitude $\hat b_{-}$,
it does not develop, that is, it remains in its initial vacuum state. Besides, if we are interested in the states of spin waves with amplitudes $\hat
b_1$ and $\hat b_2$, they can be expressed by equalities $\hat b_{1,2}=(\hat b_+\pm\hat b_-)/\sqrt2$.

In order to derive a kernel of the integral transformation $G_{ab}(t,z)$, we can take advantage of the fact that Heisenberg equations here are
identical in form to the $\Lambda$-type memory equations. Thus according to, e.g., \cite{Golubeva-2012}
\BE G_{ab}(t,z) =  \int^{T_W}_0 dt^\prime e^{-i t^\prime } \; J_0 \( \sqrt{z t^\prime} \)  \; e^{i (t - t^\prime)} J_0 \( \sqrt{z(t - t^\prime)} \) .
\L{15} \EE
As we can see, these kernels are the same for all values of $r$ in line with the fact that all equations coincide with each other in the form. $J_0$
is a zero-order Bessel function of the first kind. It is easy to see that $G_{ab}(t,z)=G^\ast_{ab}(t,z)$. Note, that we have managed the formal
coincidence of the kernels within the different configuration of the atomic excitation by choosing the mentioned above ratio between the amplitudes
of the driving fields in various configurations.

Just as we can write the signal pulse into one specific spin wave, we also can read out the excitation of the specific spin wave that determined by
the equality:
\BY
&& \hat{a}_{out}(t)= - \frac{1}{\sqrt2} \int^{L\!\!\!\!\!\!}_0 dz \; \h{ b}_r(z) \;G_{ba}(t,z) +\hat v_{out}(t).\L{16}
\EY
Here, for the case of the broadband memory under consideration, $G_{ba}(t,z)=G_{ab}(t,z)$.

The Eq. (\ref{16}) can not be interpreted as a tool to write the amplitude of the signal pulse with reference to any value of $r$. For example, if
the reading is produced from the spin wave $\hat b_+$ (under the condition $\Omega_1=\Omega_2$), then the Eq. (\ref{16}) is only suitable for $r=+$.
When $\Omega_1=-\Omega_2$ the reading is occurred from the spin wave with the index $r=-$. $ \Omega_2=0$ provides the read out from the spin wave
with $r=1$, and, $\Omega_1=0$ -- from the wave with $r=2$. Note, that Eq. (\ref{16}) also describes a situation where the writing and the reading
processes are performed in different bases. For instance, we can map the signal in the spin wave $\hat b_+$ but read out the excitation from the wave
$\hat b_1$.

Let us introduce the following function:
\BY
&&G(t,t^\prime)=\frac{1}{2} \int_0^L dz G_{ab}(t,z)G_{ba}(t^\prime,z)\L{19}
\EY
In the case of $\Lambda$-type broadband memory, this value had a meaning of the kernel of a full memory cycle for the backward retrieval of the
signal \cite{Golubeva-2012}. This meaning with minor reservations (which essence will be evident from the further discussion) survived for the tripod
atomic configuration.

The important fact is that the function of two variables, $G(t,t^\prime)$ is symmetric with respect to its permutations, i.e.
\BY
&&G(t,t^\prime)=G(t^\prime,t).\L{20}
\EY
This allows us to derive the equation for the eigenfunctions and eigenvalues of this propagator:
\BE \sqrt{\lambda_i} \;\varphi_i(t)= \int_0^{T_W}dt^\prime\; \varphi_i(t^\prime)\;G(t,t^\prime).\L{21} \EE
Here, eigenfunctions $\varphi_i(t)$ form a complete orthonormal set, i.e., they obey the relations
\BE \int_0^{T_W} dt \; \varphi_i(t)\varphi_j(t)=\delta_{ij},\qquad \sum_i\varphi_i(t)\varphi_i(t^\prime)=\delta(t-t^\prime). \L{22}\EE
%

\section{Schmidt decomposition for kernels $G_{ab}(t,z)$ and $G_{ba}(t,z)$ \L{IV}}

Schmidt decomposition is widely used for propagators in quantum memory problems since it allows to obtain the more simple and physically transparent
equations. In particular, the application of the Schmidt mode technique allows to reveal the number of independent memory channels, which are
available to use in a specific configuration. However, the well-known technique of decomposition can not be applied directly to the kernels
$G_{ab}(t,z)$ and $G_{ba}(t,z)$ in the case of the broadband memory because they do not possess the necessary properties of symmetry with respect to
the argument permutation. Here, we employ the indirect method \cite{Tikhonov-2015} using the the connection between the function $G(t,t^\prime)$ and
$G_{ab}(t,z)$ and $G_{ba}(t,z)$ via the relation (\ref{19}).

From the equation (\ref{21}), taking into account the properties of completeness and orthonormality (\ref{22}), one can get Schmidt decomposition:
\BY
&&G(t,t^\prime)=\sum_i\sqrt{\lambda_i} \;\varphi_i(t)\;\varphi_i(t^\prime).\L{23}
\EY
Let us suppose that for kernels $G_{ab}(t,z)=G_{ba}(t,z)$ one also can construct Schmidt decomposition in the form
\BY
&&G_{ab}(t,z )=G_{ba}(t,z )=\sum_i\sqrt{\mu_i} \;\varphi_i(t)\;g_i(z).\L{24}
\EY
Here, as before, $\varphi_i(t)$ is the set of orthonormal eigenfunctions of the Hermitian matrix $G(t,t^\prime)$. The values $\mu_i$ and functions
$g_i(z)$ are defined by the following requirements. First of all, we demand that the functions $g_i(z)$ also form the complete orthonormal set:
\BE \int_0^{L\!\!\!} dz \; g_i(z)g_j(z)=\delta_{ij},\qquad \sum_ig_i(z)g_i(z^\prime)=\delta(z-z^\prime). \EE
Then we substitute the decompositions (\ref{23}) and (\ref{24}) into the Eq. (\ref{19}). Applying the conditions of orthonormality, we carry out all
necessary integrations and obtain the connection between two sets of functions $g_i(z)$ and $\varphi_i(t)$ in the explicit form:
\BE \sqrt{\mu_i} \;g_i(z)=  \int_0^{T_W}dt \;\varphi_i(t)\;G_{ab}(t,z),\qquad \mu_i=\sqrt{4\lambda_i}. \EE
%

\section{Mode structure of the input signal pulse \L{V}}

As in the previous work \cite{Losev-2016}, we choose here the synchronized sub-Poissonian laser operating in CW regime as a source of nonclassical
light. The pulse with the duration of $T_W$ is cut out from the stationary radiation. This duration should be large enough to preserve the initial
 quadrature squeezing in the pulse.

Let us represent the signal pulse amplitude at the input of the memory cell as a linear superposition of Schmidt modes $\varphi_i(t)$:
\BY
&&\hat a_{in}(T_W-t)=\sum_i\hat e_{in, i}\;\varphi_i(t). \L{27}
\EY
The amplitudes of modes are determined by values $\hat e_ {in, i}$, which satisfy the canonical commutation relations,  $[\hat e_{in, i},\hat
e^\dag_{in, j}]=\delta_{ij}$. Then the inverse transformation is given by
\BY
&& \hat e_{in, i}=\int_0^{T_W}dt\; \hat a _{in}(T_W-t)\;\varphi_i(t).\L{30}
\EY
In order to determine the mode structure of the radiation one should calculate the average number of photons in every mode, i.e. the value
$\langle\hat e^\dag_{in, i}\hat e_{in, i}\rangle$. According to (\ref{30}), we get
\BY
&&\langle\hat e^\dag_{in, i}\hat e_{in, i}\rangle=\iint_0^{T_W}dt dt^\prime\; \langle\hat a^\dag_{in}(T_W-t)\hat
a_{in}(T_W-t)\rangle\;\varphi_i(t)\varphi_i(t^\prime).\L{31}
\EY
Due to the fact that the laser is synchronized, its radiation is coherent and the amplitude of the signal can be written as a sum of the large
coherent component and small fluctuations:
\BY
&&\hat a_{in}(t)=\sqrt { \o n} +\delta\hat a_{in}(t),\qquad\sqrt {\o n}\gg\delta\hat a_{in}(t).\L{28}
\EY
Here, the value $\o n$  has the meaning of the average number of photons per second.

Keeping the dominant term only under the integral in Eq. (\ref{31}), we obtain
\BY
&&\langle\hat e^\dag_{in, i}\hat e_{in, i}\rangle=\o
n\;T_W\;\varphi^2_i(\omega=0),\qquad\varphi_i(\omega)=\frac{1}{\sqrt{T_W}}\int_0^{T_W}dt\varphi_i(t)e^{i\omega t}.\L{33}
\EY
As one can see, the occupancy of Schmidt modes by photons in the input signal can be expressed by the set of values $\varphi_i(\omega=0)$. For
example, as we have shown in \cite{Golubeva-2015}, only two first Schmidt modes are occupied when $L=10$ and $T_W=5.5$ since for these parameters
each of the values $\varphi_1^2(\omega=0)$ and $\varphi_2^2(\omega=0)$ is close to $1/2$, and the others are almost zero.

Let us demonstrate now that each of occupied Schmidt mode turns out in a squeezed state. For this aim we calculate the mean square of a fluctuation
of quadrature. Of course, we suppose that the $x$-quadrature should be squeezed in Schmidt mode basis, like it was initially squeezed in the plane
wave basis for laser radiation. So, we will follow the value $\langle\delta\hat x^2_{in,i}\rangle$, having in mind that
\BY
&&\delta \hat e_{in, i}=\delta \hat x_{in, i}+i\;\delta \hat y_{in, i}.
\EY
From Eq. (\ref{30}) we obtain the equation for the normally ordered average:
\BY
&&\langle:{\delta \hat x^2_{in, i}}:\rangle=\iint_0^{T_W}dt dt^\prime\langle:\delta\hat x_{in}(T_W-t)\delta\hat
x_{in}(T_W-t^\prime):\rangle\varphi_i(t)\varphi_i(t^\prime).\L{}
\EY
Let us substitute now the explicit expression for laser generation \cite{Golubeva-2008} for the pair correlation function in integrand:
\BY
&&\kappa \langle :  \delta \hat x_{in}( t) \delta \hat x_{in}({t}^\prime) : \rangle =- \frac{1}{8} \frac{(1 - \mu)}{(1 - \mu/2)^2}\kappa (1 -
\mu/2)\; e^{\ds- \kappa (1 - \mu/2) | {t} - {t}^\prime|}.\L{34}
\EY
Here $\mu$ is a parameter of the laser synchronization. We ensure $\mu\ll 1$ to preserve the quantum properties of laser generation. Besides, for the
same purpose, we should choose the pulse duration (under its cutting out from the steady flow) long enough,  $\kappa T_W\gg1$ (where $\kappa$ is a
spectral width of the laser mode) \cite{Samburskaya-2012}. The latter requirement means that the correlation function in integrand develops much
faster than Schmidt modes $\varphi_i(t)$. So, we can make a replacement under the integral sign,
\BY
&&\kappa (1-\mu/2)e^{\ds -\kappa (1-\mu/2)(t-t^\prime) }\to\delta(t-t^\prime),
\EY
then the result of integration is given by
\BY
&&\langle:({\delta \hat x_{in, i}})^2:\rangle=-1/4,\qquad\langle({\delta \hat x_{in, i}})^2\rangle={\mu^2}/{16}\ll1.\L{36}
\EY
Thus, the $x$-quadrature of every occupied Schmidt mode is squeezed as mach as the $x$-quadrature of the laser generation.

\section{Manipulations with one squeezed signal pulse at the input of the cell \L{VI}}
\subsection{Efficiency of the spin wave excitation}
Let us start with the simplest physical situation, which has already been partially considered in \cite{Losev-2016}. We assume that the single signal
pulse in the squeezed state with the amplitude $\a_{in}(t)$ passes to the input of the tripod memory cell. As we discussed in previous Section, this
pulse is "cut out"\ from the steady light flow of the synchronized sub-Poissonian laser. The writing process of the pulse is determined by two
driving fields with Rabi frequencies $\Omega_1$ and $\Omega_2$. If only one of the frequencies, e.g. $\Omega_1$, is nonzero, then only one spin wave
with the amplitude $\b_1(z)$ is excited. However, if the atomic medium is described in terms of amplitudes $\b_\pm(z)$, then for the same driving
both spin waves $\b_+(z)$ and $\b_-(z)$ are excited equally. Vice versa, if at the writing $\Omega_1=\Omega_2$, then only one spin wave $\b_+(z)$ is
excited. This means, both waves $\b_1(z)$ and $\b_2(z)$ are excited equally. Keeping in mind the Schmidt mode decompositions (\ref{24}) and
(\ref{27}), and taking into account decomposition
\BY
&& \hat{b}_r(z)= \sum_i \hat e_{r, i}\;g_i(z), \qquad \[\hat e_{r, i},\hat e_{s, j}\]=\delta_{rs }\delta_{ij},\qquad r,s =1,2,\pm,\L{38}
\EY
it is possible to rewrite Eq. (\ref{14}) as a set of equations:
\BY
&& \hat{e}_{r, i}= -\lambda_i^{1/4} \hat e_{in, i}  +\hat v_{r, i}.\L{38}
\EY
As mentioned above, if, for example, $\Omega_1=\Omega_2$, we should apply this equation to $r=+$. Then the complete set of solutions is given by
\BY
\Omega_1=\Omega_2=\Omega/\sqrt2:&& \hat{e}_{r=+, i}= -\lambda_i^{1/4} \hat e_{in, i}  +\hat v_{+, i},\quad \hat{e}_{r=-, i}=
\hat{v}_{-, i},\nn\\
&&\hat{e}_{r=1, i}= {1}/{\sqrt2}\;\hat{e}_{r=+, i}+\hat{v}_{1, i},\quad\hat{e}_{r=2, i}= {1}/{\sqrt2}\;\hat{e}_{r=+, i}+\hat{v}_{2, i}.\L{39}
\EY
If, on the other hand, the excitation of spin waves is carried out under the condition of $\Omega_2=0$, then the complete solution is written in the
form
\BY
\Omega_1=\Omega,\; \Omega_2=0:&& \hat{e}_{r= 1, i}= -\lambda_i^{1/4} \hat e_{in, i}  +\hat v_{1, i},\qquad\hat{e}_{r=2, i}= \hat{v}_{2, i},\nn\\
&&\hat{e}_{r=+, i}= {1}/{\sqrt2}\;\hat{e}_{r=1,i}+\hat{v}_{+, i},\qquad\hat{e}_{r=-, i}= {1}/{\sqrt2}\;\hat{e}_{r=1, i}+\hat{v}_{-, i}.\L{41}
\EY
Let us estimate the efficiency of excitation of the $i$-th mode, ${\cal E}_{r,i}$, for the case $r=+$:
\BY
&& {\cal E}_{r=+,i}={\langle\hat{e}^\dag_{r=+, i}\hat{e}_{r=+, i}\rangle}/{\langle\hat e^\dag_{in, i}\hat e_{in, i}\rangle}= \sqrt{\lambda_i}.\L{}
\EY
As expected, the writing efficiency is entirely determined by the properties of the specific kernel: the modes with ${\lambda_i}$ closer to unit are
mapped better.

When we write the signal into the wave $r=+$, of course, both spin waves $\b_1$ and $\b_2$ turns out to be excited. The level of its excitations can
be obtained explicitly. Taking into account the equality $\b_\pm=(\b_1\pm\b_2)/\sqrt2$, we can see that these waves are excited equally with the
average number of excitations
\BY
&& \langle\hat{e}^\dag_{r=1, i}\hat{e}_{r=1, i}\rangle= \langle\hat{e}^\dag_{r=2, i}\hat{e}_{r=2,
i}\rangle=1/2\;\sqrt{\lambda_i}\;\langle\hat{e}^\dag_{in, i}\hat{e}_{in, i}\rangle.\L{}
\EY
Let us recall that according to Eq. (\ref{33}), the average number of photons in the input signal $\langle\hat{e}^\dag_{in, i}\hat{e}_{in, i}\rangle$
is equal to $({\o n\;T_W }\;\varphi^2_i(\omega=0))$, and also depends essentially on the mode number.

\subsection{Generation of the combined (semi-squeezed and semi-entangled) quantum state (CQS) of spin waves}

In the case of $\Omega_1=\Omega_2$,  both spin waves $\b_1$ and $\b_2$ are excited, and, in the other basis, the only spin wave $\b_+$ is excited,
but the wave $\b_-$ remains in vacuum state. It is clear intuitively for a good memory, the quantum properties of the spin wave $\b_+$ turn out to be
close to ones of the input signal, i.e. this wave is squeezed. Indeed, this conclusion is easily confirmed by Eq. (\ref{39}) for $\lambda_i\approx
1$: $\;\;\hat{e}_{r=+, i}= - \hat e_{in, i}  +\hat v_{r, i}$. Taking into account Eq. (\ref{36}), one can get
\BY
&& 4\langle:\delta\hat{x}^2_{r=+, i}:\rangle=\sqrt{\lambda_i}\;4\langle:\delta\hat x^2_{in, i}:\rangle=-
\sqrt{\lambda_i},\qquad4\langle\delta\hat{x}^2_{r=+, i}\rangle=1- \sqrt{\lambda_i}.\L{45}
\EY
Thus, for the effective memory process, statistical properties of the input pulse are well-mapped on the properties of the spin wave $\b_+$ .

In regard to the spin waves $\b_1$ and $\b_2$, each of them is only half squeezed at best:
\BY
&& 4\langle:\delta\hat{x}^2_{r=1, i}:\rangle=4\langle:\delta\hat{x}^2_{r=2, i}:\rangle= -\sqrt{\lambda_i}/2.\L{44}
\EY
Now, let us make sure that quantum properties of these spin waves are not confined to their quadrature squeezing, but they are also entangled with
each other. For this aim we can follow, for example, the Duan criterion, which can be written in the form:
\BY
&& D=\langle(\delta\hat{x}_{r=1, i}+\delta\hat{x}_{r=2, i})^2\rangle+\langle(\delta\hat{y}_{r=1, i}-\delta\hat{y}_{r=2, i})^2\rangle<1.\L{47}
\EY
To estimate this inequality let us rewrite it over the normal ordering averages and taking into account the equation (\ref{45}),
\BY
&& D=2\langle\delta\hat{x}_{r=+, i}^2\rangle+2\langle\delta\hat{y}_{r=-, i}^2\rangle=1+2\langle:\delta\hat{x}_{r=+,
i}^2:\rangle+2\langle:\delta\hat{y}_{r=-, i}^2:\rangle=1-\sqrt{\lambda_i}\;/2 .\L{}
\EY
The most effective entanglement occurs when $D\ll 1$. In our case, this variable can not be lower than $1/2$. It means, that the spin waves with
$\lambda_i\approx 1$ are really entangled but not more than on 50\%.


\subsection{Reading out of the spin wave excitations}
Let us apply the Schmidt decomposition to the Eq. (\ref{16}), which describes the reading out process. Then we obtain the following equations for the
output signal:
\BY
&& \hat e_{out,i}= -  (\lambda_i)^{1/4}\hat e_{r,i}+\hat v_{out,i},\qquad r=1,2,\pm.\L{47}
\EY
Here, $\hat e_{r,i}$ and  $\hat e_{out,i}$ are the coefficients of Schmidt decomposition for the spin waves (\ref{38}) and the retrieved signal,
respectively. The decomposition of the output retrieved signal is written in the form:
\BY
&& \a_{out}(t)=\sum_i \hat e_{out,i}\; \varphi_i(t).
\EY
As we discussed above, the choice of the amplitudes $\hat e_{r,i}$ in the Eq. (\ref{47}) depends on the reading strategy. In the case of
$(\Omega_1=\Omega_2)$, the reading is carried out from the spin wave $\b_+$, i.e. $r=+$. In the case $(\Omega_1=\Omega ,\;\Omega_2=0)$, the wave
$\b_1$ is read out, that is $r=1$, etc.

Now, for certainty, we assume that the excitation of the spin waves was provided under the condition $\Omega_1=\Omega_2$. Then the excitation is
determined by Eqs. (\ref{39}). As we can see, all spin amplitudes are excited except $\b_-$, which remains in the vacuum state. The reading can be
addressed separately to each of the excited wave. For example, if the reading (as well as the writing) is ensured by $\Omega_1=\Omega_2$, then it is
carried out from the wave $\b_+$, and the output signal is given by
\BY
&& \Omega_1=\Omega_2:\qquad\hat e_{out,i}=-(\lambda_i)^{1/4} \hat e_{r=+,i}+\hat v_{out,i}=\sqrt{\lambda_i}\;\hat e_{in,i}+\hat v_{out,i}.
\EY
From this equation one can see that the photon numbers of the input and the output signals are almost the same for the good memory, i.e. when
$\lambda_i\approx 1$, consequently, the reading is complete (almost complete). We can conclude from here that, first, the tripod memory can provide
the high efficiency and, second, quantum statistical properties of the initial signal pulse can be restored. Thus, the operation of the tripod memory
configuration is no worse than the $\Lambda$-type memory.

It is significant to note here that this memory process is similar to the passing of the squeezed signal through the Mach-Zehnder interferometer,
when the second port of the input beam splitter is not illuminated (in a vacuum state). Then, after the input beam splitter, two beams turn out in
the CQS (semi-squeezed and semi-entangled state) inside the interferometer. After the second beam splitter, we see again the pulse in the squeezed
state (and the vacuum field on the second port, which we don't care about).

Now, let us consider the reading process under the conditions $\Omega_1=\Omega,\Omega_2=0$. This ensured the reading only from the wave $\b_1$, and
the corresponding signal at the output of the cell is given by:
\BY
&&  \Omega_1=\Omega,\Omega_2=0:\qquad\hat e^{(1)}_{out,i}=-(\lambda_i)^{1/4} \hat e_{r=1,i}+\hat v_{out,i}=\sqrt{\lambda_i/2}\;\hat e_{in,i}+\hat
v_{out,i}.
\EY
One can see, the quantum statistical properties of the output signal are nearly repeat now the properties of the spin wave $\b_1$. As we remember,
the spin wave $\b_1$ was in the semi-squeezed state, consequently, the signal wave $\hat e^{(1)}_{out,i}$ also becomes semi-squeezed. Moreover,
moving away from the memory cell, the retrieved pulse remains partially entangled with the other spin wave $\b_2$, i.e. with the atomic ensemble. One
can be convinced of this again on the basis of the Duan criterion. Taking the acceptable phase of observation, so that $\hat{x}_{r=1,
i}\to-\hat{x}_{r=1, i}$ and $\hat{y}_{r=1, i}\to-\hat{y}_{r=1, i}$, we obtain
\BY
&& D=\langle(\sqrt{\lambda_i}\;\delta\hat{x}_{r=1, i}+\delta\hat{x}_{r=2, i})^2\rangle+\langle(\sqrt{\lambda_i}\;\delta\hat{y}_{r=1,
i}-\delta\hat{y}_{r=2, i})^2\rangle.\L{}
\EY
This equation coincides with the Eq. (\ref{47}) for eigenvalues $\lambda_i\approx 1$, i.e. the Duan criterion is satisfied. Thus, we can say, that
after the memory process the atomic ensemble and the retrieved signal pulse are in the CQS.

In order to read out the remaining excitation of the spin wave $\b_2$, we have to act by the driving field $\Omega_2=\Omega$ and put $\Omega_1=0$.
Then, we can obtain the second signal pulse at the output  is
\BY
&&  \Omega_2=\Omega, \Omega_1=0:\qquad\hat e^{(2)}_{out,i}=-(\lambda_i)^{1/4} \hat e_{r=2,i}+\hat v_{out,i}=\sqrt{\lambda_i/2}\;\hat e_{in,i}+\hat
v_{out,i}.
\EY
As a result, instead of the single signal pulse in squeezed state at the input of the memory cell we obtain two spatially separated pulses in CQS at
the output.

This scenario of the reading has a similarity with the passing of the signal through the beam splitter when it is mixed with the vacuum field.

In this Section we have considered as the examples three different states of the output signal, when the writing is ensured by
$\Omega_1=\Omega_2=\Omega/\sqrt2$. One can make certain that the writing under conditions $\Omega_1=\Omega,\Omega_2=0$, does not change results in
principle. As before, we have the opportunity to retrieve the single pulse with the properties close to ones of the input signal, or read out a part
of excitation and produce light-matter CQS, or, at last, read out two separated signals (at the necessary distance from each other) with the
statistics close to ones of the spin waves.


\section{Manipulation with two signal pulses at the input \L{VII}}
As we have discussed above, when $\Omega_1=\Omega_2=\Omega/\sqrt2$, the input pulse is absorbed by the resonant medium exciting the spin wave $\b_+$.
At the same time, the spin wave $\b_-$ is not involved in the interaction and remains in the vacuum state. It is possible to excite it by sending the
second signal accompanied by driving fields $\Omega_1=-\Omega_2$ to the same memory cell. The amplitudes of the spin waves are expressed then by
\BY
&& \hat{e}_{r=+, i}= -\lambda_i^{1/4} \hat e^{(1)}_{in, i}  +\hat v_{r=+, i},\quad \hat{e}_{r=-, i}= -\lambda_i^{1/4} \hat e^{(2)}_{in,
i}+\hat{v}_{r=-,i}.\L{55}
\EY
Here, $\hat e^{(1,2)}_{in, i}=\hat x^{(1,2)}_{in, i}+i\hat y^{(1,2)}_{in, i}$ are amplitudes of the first and the second signal pulse at the input of
the cell, respectively. We consider the case, when the first pulse is squeezed in $x$-quadrature, and the second -- in $y$-quadrature:
\BY
&& 4\langle: [\delta\hat {x}^{(1)}_{in, i}]^2:\rangle=4\langle: [\delta\hat {y}^{(2)}_{in, i}]^2:\rangle\; \rightarrow -1,\qquad4\langle [\delta\hat
{x}^{(1)}_{in, i}]^2\rangle=4\langle [\delta\hat {y}^{(2)}_{in, i}]^2\rangle \ll1.\L{}
\EY
Eqs. (\ref{55}) allow us to conclude, that for all modes with eigenvalues $\lambda_i\approx 1$, the spin waves are squeezed as effectively as initial
signals. Indeed,
\BY
&& 4\langle [\delta\hat {x}^{(2)}_{r=+, i}]^2\rangle=4\langle [\delta\hat {y}^{(2)}_{r=-, i}]^2\rangle=1-\sqrt{\lambda_i}.\L{}
\EY
As regards the amplitudes $\hat{e}_{r=1, i}$ and $\hat{e}_{r=2, i}$, they can be expressed via the amplitudes $\hat{e}_{r=\pm, i}$ as their sum and
difference, it results in
\BY
&& \hat{e}_{r=1, i}= \(\hat{e}_{r=+, i}+\hat{e}_{r=+, i}\)/{\sqrt2}+\hat{v}_{r=1,i}=-(\lambda_i)^{1/4}\;\(\hat e^{(1)}_{in, i}+\hat e^{(2)}_{in, i}\)/{\sqrt2}+\hat{v}_{r=1,i},\L{58}\\
&& \hat{e}_{r=2, i}=\(\hat{e}_{r=+, i}-\hat{e}_{r=+, i}\)/{\sqrt2}+\hat{v}_{r=2,i}=-(\lambda_i)^{1/4}\(\hat e^{(1)}_{in, i}-\hat e^{(2)}_{in,
i}\)/{\sqrt2}+\hat{v}_{r=1,i}.\L{59}
\EY

According to these equations, we can conclude that the quadrature fluctuations of the spin waves $\b_1$ and $\b_2$ become higher since they are
formed by the low fluctuations of one pulse and the high fluctuations of the other one. The squeezing of the spin waves vanishes but these waves turn
out to be effectively entangled (of course, for $\lambda_i\approx1$ only). Indeed, since $D$ for this case can be expressed by the following way,
\BY
&&D= 4\langle [\delta\hat {x}_{r=+, i}]^2\rangle+4\langle[\delta\hat {y}_{r=-, i}]^2\rangle=1-\sqrt{\lambda_i},\L{60}
\EY
the Duan criterion is satisfied. This result is consistent with what is obtained by mixing two fields squeezed in the orthogonal quadratures on the
symmetrical beam splitter.

Now let us consider two different scenario of the read out.
\\
\\
\emph{The first scenario:}
\\
\\
The complete read out is provided by two successive driving pulses. One of them, for example, with the Rabi frequency $\Omega_1=\Omega$ and
$\Omega_2=0$, provides the reading from the spin wave $\b_1$. The other one, with the Rabi frequency $\Omega_2=\Omega$ and $\Omega_1=0$ provides the
reading from the spin wave $\b_2$. Then the amplitudes of the signals can be given by
\BY
&& \Omega_1=\Omega, \Omega_2=0:\qquad\hat{e}^{(1)}_{out, i}= -\lambda_i^{1/4} \hat e_{r=1, i}  +\hat v_{out, i}= \lambda_i^{1/2} \(\hat e^{(1)}_{in, i}+\hat e^{(2)}_{in, i}\)/{\sqrt2}  +\hat v_{out, i},\L{61}\\
&& \Omega_2=\Omega, \Omega_1=0:\qquad\hat{e}^{(2)}_{out, i}= -\lambda_i^{1/4} \hat e_{r=2, i}  +\hat v_{out, i}= \lambda_i^{1/2} \(\hat e^{(1)}_{in,
i}-\hat e^{(2)}_{in, i}\)/{\sqrt2}  +\hat v_{out, i}.\L{62}
\EY
This implies the high level of quadrature fluctuations in both output signals because each output signal is formed by both input pulses. However,
these two output pulses are entangled. One can make sure that the Duan criterion here is satisfied also well as for spin waves, and the value $D$
becomes much smaller than the unity.

In this case the full memory cycle is also equivalent to the mixing of waves on the beam splitter, when two squeezed pulses are converted into two
entangled ones.

Besides, if the only one pulse is read out, then this pulse turns out to be entangled effectively with the spin wave (light-matter entanglement).
\\
\\
\emph{The second scenario:}
\\
\\
Let us consider two successive driving pulses so that each pulse reads out the excitations from two spin-waves $\b_1$ and $\b_2$ simultaneously. The
first pulse consists of two driving fields with $\Omega_1=\Omega_2$ and for the second one -- with $\Omega_1=-\Omega_2$. In other words, the first
pulse reads out information from the spin wave $\b_+$, and the second -- from the spin wave $\b_-$. Formally this is represented by equations
\BY
&& \Omega_1=\Omega_2:\qquad\hat{e}^{(1)}_{out, i}= -\lambda_i^{1/4} \hat e_{r=+, i}  +\hat v_{out, i}= \lambda_i^{1/2} \hat e^{(1)}_{in, i}   +\hat v_{out, i},\L{63}\\
&& \Omega_1=-\Omega_2:\qquad\hat{e}^{(2)}_{out, i}= -\lambda_i^{1/4} \hat e_{r=-, i}  +\hat v_{out, i}= \lambda_i^{1/2} \hat e^{(2)}_{in, i}   +\hat
v_{out, i}.\L{64}
\EY
From these equations one can directly arrive at a conclusion, that both pulses are statistically independent and squeezed as well as the input
signals. That is true only for Schmidt modes with $\lambda_i\approx 1$ as ever. This light transformation is naturally associated with the passing of
two squeezed beams through the Mach-Zehnder interferometer.

The results of this Section can be easily generalized on the scheme with two entangled signal pulses at the input of the memory cell. It can be
implemented by the following substitutions in all equations:
\BY
&&\hat e^{(1)}_{in, i}\to(\hat e^{(1)}_{in, i}+\hat e^{(2)}_{in, i})/\sqrt2,\qquad\hat e^{(2)}_{in, i}\to(\hat e^{(1)}_{in, i}-\hat e^{(2)}_{in,
i})/\sqrt2.
\EY
This results to the same conclusions: one of the read out scenario brings to the creation of two statistically independent squeezed pulses at the
output of the memory cell, and the other one -- to generation of two entangled signals after the storage like at the input of the cell.

\section{Conclusion \L{VIII}}

The application of the tripod atomic configuration in quantum optics provides the researcher with a number of different possibilities. First, this
system can operate as a memory cell as well as the $\Lambda$-type quantum memory. One can ensure the writing and reading procedure in that way the
statistical properties of the output signal repeat the ones of the input pulse. Herewith, the memory process can be associated with the light passing
through the Mach-Zehnder interferometer. This enables, for example, to use memory cells for interference measurements. Note that in contrast to the
article \cite{Lee-2014}, we propose an interference scheme within a single tripod configuration of the atomic ensemble.

Second, we discussed the possibility to manipulate the quantum states of signal pulses. As an example it was shown that the single signal pulse in
the squeezed state at the input of the system can be converted into two successive pulses in the combined quantum state at the output. This process
is quite comparable with what happens when light in the squeezed state is mixed with the light in the vacuum state on the beam splitter. Besides, if
the input signal is composed of two successive independent pulses which are squeezed in different quadratures, then these two pulses become entangled
at the output. It can also be compared with the mixing of such pulses on the beam splitter.

The last example of two pulses at the input of memory cell gives us another opportunity to manipulate the spin states. As we have shown, we can
provide not the complete read out of the spin excitations, but the partial one, only from one of the spin waves, leaving another wave unchanged. In
this case the atomic ensemble turns out to be entangled with the retrieved signal. Applying some acceptable procedures to the read pulse (for
example, heralded temporal photon shaping \cite{Averchenko-2016}), we are able to deform the state of a photon subsystem and thereby change the state
of the spin subsystem.

The reported study was supported by RFBR (Grants 15-02-03656а, 16-02-00180a and 16-32-00595).


\end{document}